\documentclass[aps,pre,twocolumn,superscriptaddress]{revtex4-1}
\usepackage{graphicx,color,amsmath,siunitx,hyperref}
\begin{document}

\title{Kinetic modeling of x-ray laser-driven solid Al plasmas via particle-in-cell simulation}

\author{R. Royle}
\affiliation{Department of Physics, University of Nevada, Reno, Nevada 89557, USA}
\author{Y. Sentoku}
\affiliation{Department of Physics, University of Nevada, Reno, Nevada 89557, USA}
\affiliation{Institute of Laser Engineering, Osaka University, Osaka Pref. 565-0871, Japan}
\author{R. C. Mancini}
\affiliation{Department of Physics, University of Nevada, Reno, Nevada 89557, USA}
\author{I. Paraschiv}
\affiliation{Voss Scientific, LLC, Albuquerque, New Mexico 87108, USA}
\author{T. Johzaki}
\affiliation{Graduate School of Engineering, Hiroshima University, Hiroshima Pref. 739-8527, Japan}

\date{\today}

\begin{abstract}
Solid-density plasmas driven by intense x-ray free-electron laser (XFEL) radiation are seeded by sources of non-thermal photoelectrons and Auger electrons that ionize and heat the target via collisions. Simulation codes that are commonly used to model such plasmas, such as collisional-radiative (CR) codes, typically assume a Maxwellian distribution and thus instantaneous thermalization of the source electrons. In this study, we present a detailed description and initial applications of a collisional particle-in-cell code, PICLS, that has been extended with a self-consistent radiation transport model and Monte-Carlo models for photoionization and KLL Auger ionization, enabling the fully kinetic simulation of XFEL-driven plasmas. The code is used to simulate two experiments previously performed at the Linac Coherent Light Source investigating XFEL-driven solid-density Al plasmas. It is shown that PICLS-simulated pulse transmissions using the Ecker-Kr{\"o}ll continuum-lowering model agree much better with measurements than do simulations using the Stewart-Pyatt model. Good quantitative agreement is also found between the time-dependent PICLS results and those of analogous simulations by the CR code SCFLY, which was used in the analysis of the experiments to accurately reproduce the observed K$_{\alpha}$ emissions and pulse transmissions. Finally, it is shown that the effects of the non-thermal electrons are negligible for the conditions of the particular experiments under investigation.
\end{abstract}

\maketitle

\section{Introduction \label{sec-intro}}

The Milky Way galaxy alone contains over 100 billion stars and at least as many planets \cite{Cassan2012}. The interiors of stars and giant planets exist in a state of high-energy-density (HED) plasma ($>0.1$ MJ/cc or 1 Mbar of pressure) which is divided into two broad categories. Hot dense matter (HDM) is the hot plasma found inside stars \cite{Kippenhahn2012}, and warm dense matter (WDM) is the strongly correlated plasma that exists deep within giant planets like Jupiter and Saturn \cite{Chabrier2009}. The study of HED matter is also of critical importance to inertial confinement fusion research \cite{Atzeni2004}. As the interiors of stars, planets, and imploding fusion capsules are inaccessible to direct measurement, we must rely on theoretical models to explain our observations. In order to validate our models, however, we must be able to create and diagnose sufficiently long-lived, well-characterized samples of HED matter at homogeneous temperatures and densities in the laboratory. A new generation light source, the hard x-ray free-electron laser (XFEL), has enabled the creation and measurement of such well-characterized plasmas at exactly solid density with short (sub-picosecond), intense (up to $10^{20}$ W/cm$^2$) x-ray laser pulses with keV photons that drive the plasma via sequential, single-photon, inner-shell photoionization. Energetic photoelectrons and secondary Auger electrons ionize and heat the plasma through collisional ionizations and thermalizing binary collisions.

The collisional-radiative (CR) atomic kinetics code SCFLY---a super-configuration version of FLYCHK \cite{Chung2005} that has been optimized for the simulation of solid-density XFEL-driven plasmas---has previously been used to reproduce experimentally observed K$_\alpha$ emission spectra \cite{Vinko2012,Ciricosta2012,Vinko2015,Ciricosta2016} and beam transmissions \cite{Rackstraw2015} with excellent agreement, providing insight into the plasma conditions such as space and time resolved temperatures and densities, opacities and emissivities, charge state distributions, and rates of atomic processes. Notably, SCFLY was used to aid in the direct measurement of the ionization potential depression \cite{Ciricosta2012} and collisional ionization rate \cite{Vinko2015} in solid-density aluminum plasmas. Despite the success of CR codes, they are  limited in that they typically assume a Maxwellian particle distribution and thus instantaneous thermalization of the fast photoelectrons and Auger electrons. The ionization rate and related plasma properties can depend on the details of the electron distribution since the collisional ionization cross section depends on the energy of the impacting electron. It remains to be shown to what extent the assumption of a thermalized distribution effects simulation results.

In this study, we present a detailed description and initial application of a unique simulation tool based on a two-dimensional collisional particle-in-cell code, PICLS \cite{Sentoku2008, Mishra2013, Sentoku2014}, which self-consistently solves the radiation transport (Sec.\,\ref{sec-RT}) and has been extended to enable the simulation of intense x-ray--matter interactions through the addition of Monte-Carlo models for subshell photoionization and the radiative (K$_{\alpha}$ emission) and non-radiative (KLL Auger ionization) decay processes resulting from K-shell photoionization (Sec.\,\ref{sec-photo}). We further describe the relevant models for collisional ionization and three-body recombination (Sec.\,\ref{sec-colionrec}), as well as continuum-lowering (Sec.\,\ref{sec-IPD}) to properly model strongly-correlated, solid-density plasmas. In Section \ref{sec-simulation}, PICLS is used to simulate two similar experiments performed at the Linac Coherent Light Source investigating XFEL-driven solid-density aluminum plasmas \cite{Vinko2012} in an effort to benchmark the code. The simulated transmissions using two widely-used continuum-lowering models are compared directly to experimental measurements and SCFLY calculations, and the time-dependent results are compared in detail with those of SCFLY. As an initial application, the code is used to determine the effect of the non-Maxwellian electron distribution on the ionization rate and related plasma properties. The results are summarized and future plans are discussed in the final section.

\section{Radiation transport in PICLS \label{sec-RT}}

The radiation transport model that has recently been implemented in PICLS to enable the simulation of kinetic, radiative plasmas solves the transport equation for specific intensity $I(\mathbf{r},\Omega,\nu,t)$ [erg/cm$^2$/sr/Hz] \cite{Mihalas1978},
\begin{equation}
\left ( \frac{1}{c} \frac{\partial}{\partial t} + \mathbf{n} \cdot \nabla \right ) I= \eta - \chi I \,,
\label{RT}
\end{equation}
where $\eta(\mathbf{r},\nu,t)$ is the emissivity [erg/cm$^3$/s/sr/Hz], $\chi(\mathbf{r},\nu,t)$ is the opacity [1/cm], $\Omega(\theta,\phi)$ is solid angle [sr], and $\nu$ is the radiation frequency [Hz]. The unit vector $\mathbf{n}=(\cos\theta, \,\sin\theta\cos\phi, \,\sin\theta\sin\phi)$ lies along the ray direction. Equation (\ref{RT}) is solved by the constrained interpolation profile (CIP) scheme\,\cite{Yabe1991}, in which the intensity profile is solved together with its derivative in order to reduce numerical diffusion and maintain 3rd-order spatial accuracy. The intensity is discretized in photon energy $h\nu$ and solved for each photon energy bin $h\nu_i$ using the multi-group method, in which several ranges or groups can be defined with different bin-densities so that higher resolutions can be used in regions of interest in order to, for example, resolve a radiation source or capture spectral features of radiative bound-bound transitions. Using the discrete ordinate method \cite{Lee1962}, the intensity is further discretized in solid angle $\Omega$ and solved in each direction $\Omega_i$ in the upper hemisphere while the lower hemisphere is assumed symmetric to reduce computational cost. 

Because the radiation transport calculation can easily become more expensive than the PIC calculation itself---both in terms of processing time and memory usage since Eq.\,(\ref{RT}) must be solved for every $h\nu_i$, $\Omega_i$, cell, and timestep---it is generally performed using a cell and timestep (rad-cell and $\Delta t_{\text{rad}}$) 5 to 10 times coarser than the PIC-cell and $\Delta t_{\text{PIC}}$. Thus each 2D rad-cell would contain 25 to 100 PIC-cells, and the $\Delta t_{\text{rad}}$ would be performed once every 5 to 10 $\Delta t_{\text{PIC}}$. It is important that the chosen rad-cell and $\Delta t_{\text{rad}}$ are small enough to capture all important spatial gradients and temporal phenomena of interest. With application to XFEL-driven plasmas, the rad-cell should be small enough to resolve gradients in the x-ray intensity, ion charge, and electron energy density, and the $\Delta t_{\text{rad}}$ should be much smaller than the average hollow-atom decay lifetime ($\sim$1 fs).

\begin{figure}[htbp]
\includegraphics[width=8.0cm]{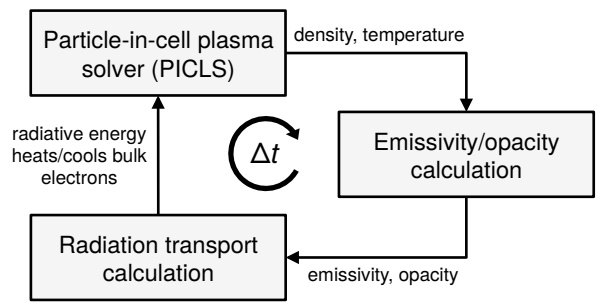}
\caption{Schematic of the computation cycle performed every $\Delta t_{\text{rad}}$ in each rad-cell.}
\label{PIC-RT}
\end{figure}

Figure\,\ref{PIC-RT} illustrates how the PIC model communicates with the radiation transport model over the course of one $\Delta t_{\text{rad}}$. Within each 2D rad-cell ($i,j$), the PIC plasma solver determines the average density, temperature, and ion charge state, from which the emissivity and opacity are determined as functions of $h\nu$. Equation\,(\ref{RT}) is then solved for $I_{ij}(\Omega,\nu)$ for each $\Omega$ and $h\nu$ value, and the total change in radiative energy is calculated as
\begin{equation}
dE_{ij}=\frac{1}{c}\iint_{\text{all}} \frac{\partial I_{ij}(\Omega,\nu,t)}{\partial t}\,d\Omega\,d\nu\,. 
\label{dE_rad}
\end{equation}
Energy transfer between the radiation field and the plasma is achieved by uniformly heating the free bulk electrons in the rad-cell if $dE_{ij}<0$ or uniformly cooling them if $dE_{ij}>0$, while respectively adding $dE_{ij}$ to or removing it from the local radiation field. 

In order to solve Eq.\,(\ref{RT}), the emissivity and opacity of the plasma must be determined. This is accomplished in PICLS in two ways. First, since photoionization (bound-free opacity) is generally the dominant absorption mechanism of hard x-rays with $h\nu$ less than a few tens of keV, an accurate and self-consistent Monte-Carlo photoionization model has been developed and is discussed in detail in Sec.\,\ref{sec-photo}. The bound-bound, bound-free, and free-free emissivity and the bound-bound and free-free opacity are determined by interpolation within tables of emissivity $\eta(n_i,T_e,h\nu)$ and opacity $\chi(n_i,T_e,h\nu)$ which are precalculated with a CR code (e.g., FLYCHK \cite{Chung2005}).

The temperature of the thermalized bulk free electrons is estimated by assuming an ideal gas as $k_{\text{B}}T_e=(2/3)E_{\text{ave}}$, where $E_{\text{ave}}$ is the cell-average kinetic energy of the bulk electrons. An electron is considered to be part of the bulk population if its kinetic energy $E<E_{\text{bulk}}$, where $E_{\text{bulk}}$ is a threshold energy value chosen as a simulation parameter. In simulations of XFEL-driven HED plasmas, $E_{\text{bulk}}$ is typically set to include all free electrons since the non-thermal electrons are usually less than a few keV. Thus if the electron distribution contains a significant population of non-thermal electrons with $E<E_{\text{bulk}}$, the calculated $T_e$ represents the temperature of a thermalized distribution with equal total energy.

\section{Photoionization and Auger decay \label{sec-photo}}

With the exception of resonant absorption from certain bound-bound transitions at specific energies, inner-shell photoionization is the dominant absorption mechanism of keV x-rays in matter. When a core K-shell electron is photoionized, the so-called hollow atom exists in an excited state with a core vacancy for a short time before decaying by either the emission of a characteristic photon (K$_{\alpha}$ or K$_{\beta}$) or the ejection of an energetic Auger electron. In the context of simulating XFEL-driven plasmas, it is important to accurately model these process since the plasma is both created and heated by the energetic photoelectrons and Auger electrons, and an abundance of hollow atoms can cause a reduced absorption of the x-rays.

Monte-Carlo-based algorithms are generally the most accurate way of including atomic physics processes in a PIC code, where cross sections or rates are used to determine the probability of an event and a random number generator decides the outcome. For example, in PICLS, the models for binary collisions, collisional (impact) ionization, and field ionization are based on this approach \cite {Sentoku2008, Mishra2013, Kato1998}. In this section, we describe the Monte-Carlo models for photoionization and hollow-atom decay including KLL Auger ionization that have been implemented in PICLS to enable the simulation of XFEL-driven plasmas. The Los Alamos suite of relativistic atomic physics codes \cite{Fontes2015} was used to calculate all photoionization cross sections, Auger and x-ray decay rates, and ionization potentials.

\begin{figure}[htbp]
\includegraphics[width=8.5cm]{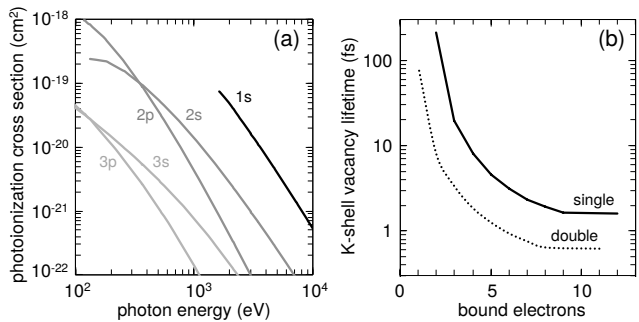}
\caption{(a) Single-electron photoionization cross sections for a neutral aluminum atom as a function of photon energy. (b) Average decay time of a hollow aluminum atom with one or two K-shell vacancies as a function of the number of remaining bound electrons. \cite{Fontes2015}} 
\label{photo-auger}
\end{figure}

Tables of subshell photoionization cross sections are prepared as a function of $h\nu$ for each atomic species as shown for aluminum in Fig.\,\,\ref{photo-auger}(a). The cross section for K-shell photoionization is at least an order of magnitude larger than those of higher orbitals (assuming $h\nu$ is larger than the K-edge energy). Neutral-atom cross sections are used since the dominant inner-shell cross sections do not vary significantly with ion charge. As an ion is further ionized, the binding energies of the remaining electrons increase due to a reduced nuclear screening, causing a shift of the absorption edges toward higher energies. This effect is accounted for by utilizing tables of subshell binding energies as a function of ion charge.

For a given configuration-average subshell $s$ (e.g., 1s, 2s, 2p, ...) of a given ion in the simulation, the probability of photoionization $P_s$ occurring within the $\Delta t_{\text{rad}}$ interval is calculated by first summing the probability for each of the $N_{h\nu}$ photon energy bins used in the simulation as
\begin{equation}
P_s=c \Delta t N_{\text{b}}\sum_{i=1}^{N_{h\nu}} n_{h\nu}(h\nu_i) \sigma_s(h\nu_i)\,,
\label{pi-prob-tot}
\end{equation}
where $n_{h\nu}$ is the local photon number density, and $\sigma_s$ is the subshell photoionization cross section. The number of bound electrons $N_{\text{b}}$ in the subshell is determined by assuming a lowest-energy configuration of the ion. A random number $0<r<1$ is generated, and if $r<P_s$, photoionization occurs, in which case another random number $0<r<1$ is generated to determine the energy of the photon responsible for the ionization. The probability for each $h\nu_i$ is accumulated (starting from $i=0$) until it surpasses $rP_s$, and the corresponding $h\nu_i$ is selected for the ionizing photon. A photoelectron is created with energy equal to the ionizing photon's energy minus the subshell binding energy, and an equal amount of energy is removed from the given energy bin of the radiation field in the rad-cell. The initial velocity of the photoelectron is randomly oriented, which is justified if its energy is low (i.e. the x-ray photon energy is not too far above the absorption edge) and the collision frequency of the plasma is high (as for solid-density plasmas) so that the photoelectron does not venture far before thermalizing. If the x-ray energy density of the ionizing photon's energy bin is not sufficient to produce one macro-photoelectron---which in a PIC code represents a large number $N_p$ of real photoelectrons---then a partial ionization is performed, creating a macro-photoelectron with fractional weight representing a number of photoelectrons less than $N_p$ and removing a fractional charge from the ion.

The excited hollow-atom state resulting from a K-shell photoionization can persist for several femtoseconds before decaying, which can become comparable to the duration of short XFEL pulses. If a significant fraction of the ions contain K-shell vacancies, the K-shell photoionization rate can be reduced and thus the decay rate can affect the overall absorption of the laser pulse. This is especially true for highly ionized atoms which also have L-shell vacancies since the K-vacancy decay rate decreases rapidly with the number of missing L-shell electrons. 

A K-vacancy lifetime model has been introduced in PICLS to account for the reduced x-ray absorption of hollow atoms. For each ion, the number of K-vacancies (0, 1, or 2) is recorded. No other electronic configuration information is stored for the ion except for its charge state $Z_i$, and so the super-configuration (SC) average K-vacancy lifetime is calculated as 
$\langle\tau_{\text{K}}\rangle_{\text{SC}}=[\langle\Gamma_a\rangle_{\text{SC}}+\langle\Gamma_x\rangle_{\text{SC}}]^{-1}$, 
where $\langle\Gamma_a\rangle_{\text{SC}}$ and $\langle\Gamma_x\rangle_{\text{SC}}$ are the SC-average Auger and x-ray decay rates for charge state $Z_i$. A super-configuration includes all transitions with initial configuration corresponding to $Z_i$.

Figure \ref{photo-auger}(b) shows the average K-vacancy lifetime for a singly and doubly hollow aluminum ion as a function of the number of remaining bound electrons $N_{\text{b}}=Z-Z_i$. 
As the L-shell becomes ionized, $\langle\tau_{\text{K}}\rangle_{\text{SC}}$ can become several femtoseconds higher than initially for the neutral atom. Since $\langle\tau_{\text{K}}\rangle_{\text{SC}}$ is approximately constant if there are no L-shell vacancies, we calculate $\langle\tau_{\text{K}}\rangle_{\text{SC}}$ only for charge states with $N_{\text{b}}<10$ (with a K-vacancy), i.e. with only electrons in the K and L shells. For charge states with $N_{\text{b}}\ge10$, $\langle\tau_{\text{K}}\rangle_{\text{SC}}$ is linearly interpolated between the value for an ion with $N_{\text{b}}=10$ and the neutral atom value with $N_{\text{b}}=Z$.

In order to determine $\langle\Gamma_a\rangle_{\text{SC}}$ and $\langle\Gamma_x\rangle_{\text{SC}}$ for a given $Z_i$, we first calculate the configuration-average Auger and x-ray decay rates $\langle\Gamma_a\rangle$ and $\langle\Gamma_x\rangle$ for every transition in which the inital state configuration has a charge of $Z_i$ and one or two K-shell vacancies. All excited configurations within the K and L shells are included. The SC-average rates are then determined by averaging the configuration-average rates weighted according to an assumed fractional population distribution proportional to the number of fine-structure substates included in each configuration.

The probability that a hollow atom will decay within the interval $\Delta t_{\text{rad}}$ is $P_{\text{decay}}=\Delta t_{\text{rad}} / \tau_{\text{K}}$ (it is important that $\Delta t_{\text{rad}} \ll \tau_{\text{K}} \sim 1$ fs). A random number $0<r<1$ is generated, and if $r<P_{\text{decay}}$, relaxation of the excited state occurs and a K-shell vacancy is filled by an electron from the L-shell. 

The final step in modeling the decay process is to determine whether the decay path is radiative or non-radiative. KLL Auger ionization (non-radiative) is the dominant path for low-$Z$ elements (e.g., $\sim$96\% probability for aluminum). The probability of Auger ionization occurring over the radiative pathway is $1-\langle\omega_{\text{K}}\rangle_{\text{SC}}$, where $\langle\omega_{\text{K}}\rangle_{\text{SC}} = \langle\Gamma_x\rangle_{\text{SC}}/[\langle\Gamma_a\rangle_{\text{SC}}+\langle\Gamma_x\rangle_{\text{SC}}]$ is the SC-average fluorescence yield of the ion. Again, a random number $0<r<1$ is generated, and if $r < P_{\text{Auger}}$, the ion decays by Auger ionization, otherwise decay by emission of a K$_\alpha$ photon occurs. In the latter case, the K$_\alpha$ photon energy is simply added to the radiation field. If Auger ionization occurs, a randomly-oriented Auger electron is created at the ion's location, and its energy is calculated as the K$_\alpha$ photon energy minus the average binding energy of the 2s and 2p subshells weighted by the number of electrons in each. Currently, only KLL Auger ionization is accounted for and the model is therefore only appropriate for low-$Z$ or mid-$Z$ targets in which higher-order Auger processes such as LMM Auger ionization and electron shake-off are not significant.

\section{Collisional ionization and three-body recombination \label{sec-colionrec}}

\begin{figure}[htbp]
\includegraphics[width=8.5cm]{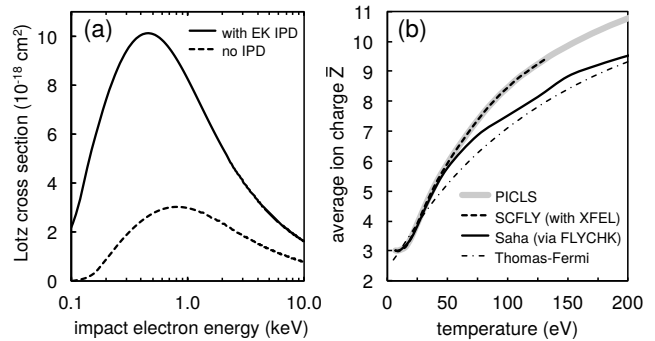}
\caption{(a) Collisional ionization cross section for an Al$^{3+}$ ion as a function of impact electron energy, both with and without the effect of continuum-lowering in solid Al with $\bar{Z}=3$. (b) EOS ($\bar{Z}(T_e)$) in solid Al as determined by the Thomas-Fermi model, FLYCHK (Saha/Boltzmann model), and SCFLY with an XFEL-driven plasma. The SCFLY EOS is assumed in the PICLS recombination model.}
\label{fig-colionrec}
\end{figure}

K-shell photoionization by an x-ray laser pulse drives the plasma by creating energetic electrons in the target at two or more distinct energies. These non-thermal photoelectrons and Auger electrons ionize and heat the target via collisional ionizations and thermalizing binary collisions. Thermalization in PICLS occurs through electron-ion and electron-electron Monte-Carlo binary collisions including the effects of collision with partially ionized atoms in both the HDM and WDM regimes \cite{Sentoku2008, Mishra2013}. A Monte-Carlo model for collisional impact ionization appropriate for non-LTE plasmas is used in PICLS that is based on the cross section derived by Lotz \cite{Lotz1970},
\begin{equation}
\sigma_{\text{ci}} = \sum_{i=1}^{N_s} a_i N_i \frac{\ln{(E/P_i)}}{EP_i}\left( 1-b_i e^{-c_i(E/P_i-1)} \right)\,,
\label{Lotz}
\end{equation}
where $E$ is the energy of the impact electron, $P_i$ and $N_i$ are the binding energy of and number of electrons in the $i$-th subshell, respectively, and $a_i$, $b_i$, and $c_i$ are individual constants determined both theoretically and experimentally which are tabulated in the reference. The sum is over the $N_s$ occupied subshells, and the contribution from subshell $i$ is zero if $E<P_i$. Fig.\,\,\ref{fig-colionrec}(a) shows $\sigma_{\text{ci}}(E)$ for an Al$^{3+}$ ion both with and without corrections to the binding energies due to continuum-lowering using the Ecker-Kr{\"o}ll model (Sec. \ref{sec-IPD}) in solid Al with average ion charge $\bar{Z}=3$. The lowered potentials experienced by ions in dense, strongly-correlated systems significantly increases $\sigma_{\text{ci}}$, causing an increased collisional ionization rate. Since $\sigma_{\text{ci}}$ rapidly decreases for large $E$, the ionization rate can be reduced for highly non-thermal electron distributions compared to a thermalized distribution with equal energy. 

In order to reduce computation times, electrons and ions are not individually paired. Instead, $\sigma_{\text{ci}}$ is calculated between each free electron and the average cell ion. A cell-average ionization probability is found to determine whether or not ionization occurs for a given ion in a Monte-Carlo fashion as described previously. When ionization does occur, a valence electron becomes a free electron at the ion's location with zero momentum. To ensure energy conservation within the cell, the free electron kinetic energies are reduced by a constant factor and a total amount equal to the sum of the ionization energies.

It is not sufficient to include only collisional ionization when simulating solid-density XFEL-driven plasmas. The inverse process of three-body recombination must also be included since the plasma temperature and density depend on the balance between the two processes. The recombination algorithm currently used in PICLS does not take a probabilistic approach using interaction cross sections as the impact ionization model does. Instead, a simplified approach is taken in which a chosen equation of state (EOS), $\bar{Z}_{\text{EOS}}(n_i,T_e)$, is used to constrain the cell-average ion charge $\bar{Z}_{\text{cell}}$. For each $\Delta t_{\text{PIC}}$, if $\bar{Z}_{\text{cell}} > \bar{Z}_{\text{EOS}}$, recombination is performed by randomly recombining a fraction of the ions with randomly selected electrons such that $\bar{Z}_{\text{cell}} \approx \bar{Z}_{\text{EOS}}$ afterwards. Energy in the cell is conserved by uniformly adding the sum of the kinetic energies and binding energies of all recombined electrons to the remaining bulk free electrons. 

It is important to choose an EOS which accurately reflects the problem under consideration. Figure \ref{fig-colionrec}\,(b) shows $\bar{Z}_{\text{EOS}}(T_e)$ for solid Al calculated by several models. The Thomas-Fermi model \cite{Salzmann1998} and the Saha/Boltzmann model (as calculated by FLYCHK \cite{Chung2005}) yield fairly similar results for $T_e<200$ eV. However, these steady-state models do not account for the additional ionization that occurs from a driving photoionization mechanism. We therefore use the EOS calculated by SCFLY \cite{Ciricosta2016} in which the plasma is driven by an x-ray laser pulse with photon energies near the Al K-edge as shown in the figure. 

There are several limitations to this technique. Most importantly, we are enforcing a chosen EOS instead of predicting it from a more fundamental standpoint. The model also does not properly account for the effects of non-thermal electrons, the presence of which can cause an overestimated $T_e$ for the EOS, and the reduced recombination cross section for high-energy electrons is ignored. Additionally, the recombination rate loses accuracy during heating and can experience non-physical oscillations (see Fig.\,\,\ref{CSD-rates}(d)) since recombination only occurs when the condition $\bar{Z}_{\text{cell}} > \bar{Z}_{\text{EOS}}$ is satisfied, and no limits are placed on the instantaneous rate, which can be overestimated. If $\bar{Z}_{\text{cell}} < \bar{Z}_{\text{EOS}}$, recombination is switched off and the rate is zero. Despite these limitations, this simple model works well enough to enable the simulation of solid-density XFEL-driven plasmas with reasonable accuracy and meaningful results, though future efforts will be directed towards the development of a recombination model based on interaction cross sections that is analogous and complimentary to the impact ionization model.

\section{Continuum-lowering \label{sec-IPD}}

All of the atomic processes discussed so far depend on ionization potentials, which can be significantly reduced through interaction with the surrounding fields in a dense, strongly-correlated plasma relative to those of an isolated ion, significantly altering the charge state distribution of the plasma. Both of the widely used Stewart-Pyatt (SP) and Ecker-Kr{\"o}ll (EK) ionization potential depression (IPD) models have been implemented in PICLS. Both models reduce to the Debye-H{\"u}ckel (DH) theory \cite{Griem1997} in the limit of low density and high $T_e$, but they give very different predictions at solid densities.
The 1963 EK IPD model \cite{Ecker1963} takes the form of two limiting cases. If the total particle density $n=n_e+n_i$ is greater than the critical density $n_{\text{crit}}=(3/4\pi)(4\pi\epsilon_0 k_{\text{B}}T_e/(Z_i+1)^2 e^2)^3$, as is typically the case at solid density, then the energy shift is given by
\begin{equation}
\Delta E_{\text{EK}} = C\frac{(Z_i+1)e^2}{4\pi\epsilon_0 r_{\text{EK}}}\,,
\label{IPD-EK}
\end{equation}
where $r_{\text{EK}}=(3/4\pi n)^{1/3}$ and the constant $C=1$ following the arguments in Ref. \cite{Preston2013}. The 1966 SP IPD model \cite{Stewart1966} interpolates between the results of the DH model in the limit of low density and high $T_e$ and the average atom ion-sphere (IS) model in the limit of high $n_e$. The energy shift is given by
\begin{equation}
\Delta E_{\text{SP}} = \frac{k_{\text{B}}T}{2(z^{\ast}+1)} \left( [3(z^{\ast}+1)K+1]^{2/3} - 1 \right)\,,
\label{IPD-SP}
\end{equation}
where $K=(Z_i+1) e^2/4 \pi \epsilon_0 \lambda_{\text{D}} k_{\text{B}}T$, and the Debye radius $\lambda_{\text{D}}=(\epsilon_0 k_{\text{B}} / e^2 n_e(\bar{Z}T_e/T_i+1))^{1/2}$, where it is assumed that $T_i=T_e=T$. The parameter $z^{\ast}=\langle Z_i^2 \rangle / \langle Z_i \rangle \approx \bar{Z}$ defines the ionization degree of the plasma. The details of and differences between the two models have recently been discussed in much greater detail elsewhere (see for example Ref.\,\,\cite{Crowley2014}). 

\begin{figure}[htbp]
\includegraphics[width=8.5cm]{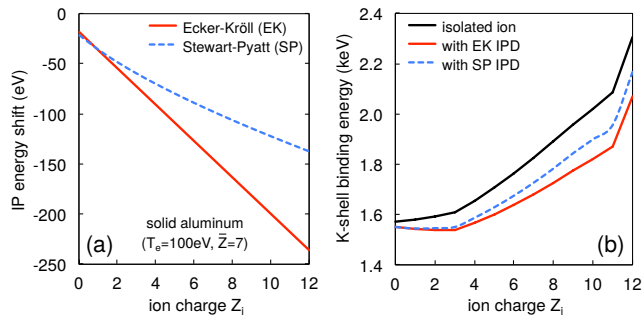}
\caption{(Color online) (a) Energy shift of ionization potentials as a function of ion charge predicted by the EK and SP models for a 100 eV, solid-density Al plasma with $\bar{Z}=7$ and corresponding $n_e$, and (b) the effect on the Al K-edge energy.}
\label{IPD}
\end{figure}

The energy shift $\Delta E$ is calculated for each ion in the simulation and subtracted from the binding energy of each suborbital. Fig.\,\,\ref{IPD}(a) demonstrates the $\Delta E$ predicted by both the EK and SP models as a function of ion charge for a 100 eV, solid-density Al plasma with $\bar{Z}=7$ and corresponding $n_e$, and Fig.\,\,\ref{IPD}(b) shows the effect of $\Delta E$ on the Al K-edge energy for the same conditions. At higher charge states, $|\Delta E_{\text{EK}}|$ can be much larger than $|\Delta E_{\text{SP}}|$.

Analysis of the results of the particular XFEL experiments under consideration in this study has shown a much better agreement with the EK model than with the SP model \cite{Ciricosta2012}, though it should be noted that another experiment using the high-power Orion laser to generate a plasma with higher density and temperature found better agreement with the SP model \cite{Hoarty2013}. Recent density functional theory (DFT) calculations have also yielded  similar results to the EK model \cite{Vinko2014}. We therefore use the EK IPD model in the PICLS simulations discussed in the following section unless otherwise specified.

For plasmas at solid density, the IPD energy shift can be larger than the isolated-atom binding energies of some of the weakly bound valence electrons, causing them to become ionized. This so-called pressure ionization is not currently accounted for in PICLS. Instead, the initial charge state of the ions is chosen to approximate the degree to which pressure ionization initially occurs. For example, in solid aluminum the three M-shell electrons are ionized due to IPD and so an initial charge state of Al$^{3+}$ is chosen.

\section{Simulations of x-ray laser-driven aluminum plasmas \label{sec-simulation}}

In this section we present results of PICLS simulations compared with the results of two similar experiments performed at the Linac Coherent Light Source (LCLS) investigating solid-density HED plasmas driven by K-shell photoionization in which thin aluminum foil targets were irradiated by short, intense XFEL pulses with photon energies at and above the cold Al K-edge. We also compare our results to those of the related SCFLY simulations performed in the analysis of the experiments. We then further investigate the effects of the non-thermal electrons on the electron and atomic kinetics of the plasma.

In PIC simulations of optical laser--matter interactions, the cell and timestep size are usually dictated by the need to resolve the laser wavelength and frequency since absorption occurs through various modes of energy coupling between the plasma electrons and the electromagnetic field of the laser. When simulating x-ray absorption, however, the angstrom-scale fields luckily need not be resolved since absorption by photoionization depends only on intensity. The spatial grids need only resolve gradients in the x-ray intensity and resulting plasma properties. In the following PICLS simulations, a rad-grid with a resolution of 30 rad-cells/\SI{}{\micro\meter} is used as it is sufficient to resolve the 7--9 \SI{}{\micro\meter}$^2$ x-ray spots. The PIC grid is 5 times finer with a resolution of 150 PIC-cells/\SI{}{\micro\meter}. The timesteps need only be small enough to provide a good statistical representation of the probabilistic atomic process with the highest rate and is chosen such that the probability of an event occurring during the timestep is much less than one. The $\Delta t_{\text{rad}}$ is chosen to be 0.1 fs, which is over an order of magnitude smaller than the average hollow-atom decay lifetime ($>$1 fs), and the $\Delta t_{\text{PIC}}$ is 5 times smaller at 0.02 fs, which is much smaller than the average time between collisional ionizations at the peak rate ($>$1 fs). The simulations have 16 ions/PIC-cell (up to 208 free electrons), and 400 ions/rad-cell (up to 5200 electrons), which is sufficient to provide good statistical representations of all atomic processes. This was verified by comparing results with test simulations using fewer particles. The simulation box size is \SI{1.6}{\micro\meter} $\times$ \SI{4}{\micro\meter}, with \SI{0.1}{\micro\meter} of vacuum on either side of the \SI{1.4}{\micro\meter} Al target. Absorbing boundary conditions are used, though the choice of boundary conditions are of little concern since the plasma is confined within the target laterally by induced sheath fields, and the keV electrons do not travel far in solid matter before thermalizing. 

\subsection{$h\nu$-dependent saturable absorption in Al}

As an initial test of the x-ray photoabsorption processes in PICLS, we simulate a recent LCLS experiment by Rackstraw \textit{et al.}\,\,\cite{Rackstraw2015} in which the saturable absorption of intense XFEL pulses with photon energies $h\nu$ scanned across the cold Al K-edge was measured. PICLS-simulated transmissions using both the Ecker-Kr{\"o}ll and Stewart-Pyatt continuum-lowering models are compared directly to the measured and SCFLY-simulated transmissions from Ref.\,\,\cite{Rackstraw2015}.

In the experiment, a \SI{1}{\micro\meter} aluminum foil target was irradiated at 45$^{\circ}$ from normal by XFEL pulses resulting from 100 fs electron bunches, giving an effective path length of \SI{1.4}{\micro\meter} and approximately 60 fs \cite{Dusterer2011} x-ray pulses. Photon energies were sampled in the range of 1540---1870 eV, from just below to well above the cold Al K-edge at $\sim$1560 eV. The PICLS simulations consist of a \SI{1.4}{\micro\meter} solid Al target and a normally incident x-ray pulse with \SI{7}{\micro\meter}$^2$ ($1/e$) Gaussian radial intensity profile and 65 fs flat-top temporal profile. The experimental pulse energies were $\sim$2 mJ before passing through focusing optics, and an on-target energy of 0.50--0.60 mJ was used for the simulations (depending on $h\nu$) assuming a 25--30\% beamline transmission (reported values are 27--34\%) and a constant $2 \times 10^{12}$ photons per pulse. The resulting peak intensities are just over $10^{17}$ W/cm$^2$. The SCFLY simulations used a slightly higher on-target energy of 0.8 mJ (40\% beamline transmission) and a longer x-ray pulse duration of 100 fs, though it has been verified here and elsewhere \cite{Ciricosta2016} that the total absorption for these conditions depends only on the total x-ray fluence and not on the specific pulse shape.

\begin{figure}[htbp]
\includegraphics[width=8.5cm]{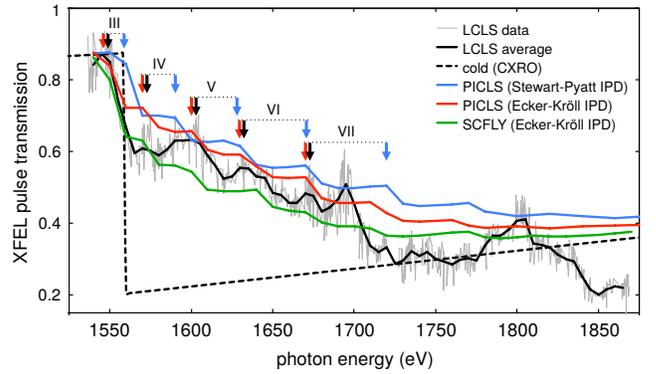}
\caption{(Color online) $h\nu$-dependent XFEL pulse transmissions in solid Al from a recent LCLS experiment and SCFLY simulations \cite{Rackstraw2015} as well as from PICLS simulations using both the EK and SP IPD models. The dashed line shows the cold Al transmission values from the CXRO database \cite{Henke1993,CXRO}. The arrows indicate approximate K-edge energies of Al ions of increasing charge states, denoted by Roman numerals.}
\label{transmission}
\end{figure}

The measured and simulated transmissions as well as the cold Al transmissions from the CXRO online database \cite{Henke1993,CXRO} are shown in Fig.\,\,\ref{transmission} versus $h\nu$. The solid black curve shows experimental trend to help guide the eye. As an ion is ionized to higher charge states, the binding energies of the remaining electrons increase due to reduced screening of the nuclear charge. If $h\nu$ is only slightly above the initial cold K-edge, the K-shell binding energy can eventually surpass $h\nu$ preventing further absorption by K-shell photoionization. Collisional ionization and L-shell photoionization also contribute in depleting the number of ions with charge states low enough to allow K-shell photoionization. If the fluence of the XFEL pulse is high enough, the absorbing ion population can become depleted before the end of the pulse causing saturation of the absorption. Thus the initial decrease in transmission just beyond the cold K-edge is not as large as predicted by the cold transmission curve as seen in the figure. As $h\nu$ increases, the pulse transmission gradually decreases, approaching the cold transmission value in small steps located at the K-edges of ions of increasing charge state. This feature provides a precise view of the K-edge energies and thus of the degree of ionization potential depression (IPD) that occurs in the dense plasmas.

We have performed PICLS simulations using both the Ecker-Kr{\"o}ll (EK) and Stewart-Pyatt (SP) IPD models, which predict increasingly different values of the IP energy shift with increasing ionization. The arrows in Fig.\,\,\ref{transmission} indicate the approximate K-edge energies observed in the experimental data (black) and PICLS simulations using the EK (red) and SP (blue) IPD models for increasing charge states (pressure ionization of the M-shell electrons results in an initial charge state of Al$^{3+}$). The K-edge energies predicted by the EK model agree very well with those observed in the experiment, while those predicted by the SP model are far too high. This is in agreement with the results of SCFLY-simulated K$_{\alpha}$ emission spectra in a separate study under similar conditions \cite{Ciricosta2012}. The EK IPD model was also used in the SCFLY simulations that produced the transmissions shown in Fig.\,\,\ref{transmission}, and the observed K-edge energies likewise agree well with the experimental values.

The PICLS-simulated transmissions tend to be slightly higher than the experimental values, especially just beyond each K-edge where the absorption is presumably enhanced by strong resonant bound-bound transitions. PICLS does not yet include a representation of excited atomic states beyond a record of K-shell vacancies and thus photoexcitation is currently neglected. In contrast, SCFLY tends to overestimate the absorption even though the pulse energy used was slightly higher than the nominal experimental value (the fractional absorption decreases as the pulse energy increases). It should be noted that there is a higher degree of uncertainty in the experimental transmissions above 1670 eV where the cold Al attenuating filters used exhibit modulations in their absorption coefficient due to x-ray absorption fine structure (EXAFS).

\subsection{Investigating time-dependent electron and atomic kinetics}

To further examine the performance of PICLS applied to the modeling of XFEL-driven plasmas, we simulate the conditions of a similar experiment performed by Vinko \textit{et al.}\,\,\cite{Vinko2012} again at the LCLS, which was the first experiment to investigate solid-density HED plasmas driven by K-shell photoionization. The focus here was to measure the time-integrated K$_\alpha$ emission spectra of the plasma rather than the pulse transmissions. The CR code SCFLY was again used to model the experiment and was able to reproduce the observed K$_{\alpha}$ spectra with excellent agreement, aiding in the direct measurement of the ionization potential depression \cite{Ciricosta2012} and collisional ionization rate \cite{Vinko2015} of the plasma. Although PICLS is capable of generating space and time resolved K$_{\alpha}$ emission data, the accuracy and quality of the data in this context has not yet been scrutinized, and so PICLS is currently unable to produce spectroscopic-quality K$_{\alpha}$ emission spectra suitable for direct comparison with experimental measurements. We therefore focus here on a comparison between the simulation results of PICLS and SCFLY.

\begin{figure}[htbp]
\includegraphics[width=8.5cm]{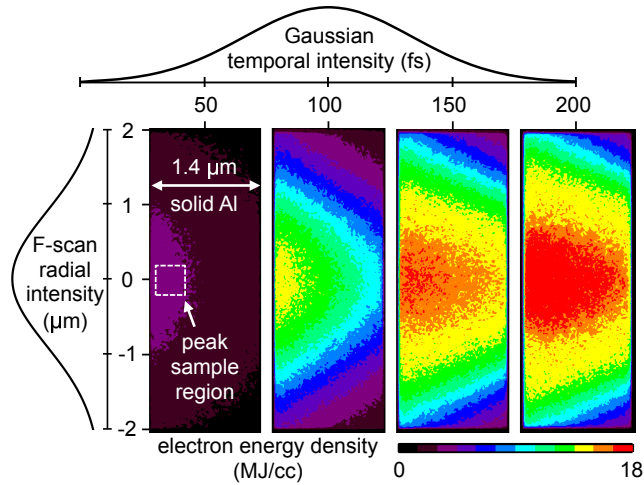}
\caption{(Color online) Electron energy density at different times for an 80 fs FWHM Gaussian x-ray pulse with 1700 eV photons and a peak intensity of $1.36\times 10^{17}$ W/cm$^2$. The radial intensity profile is taken from the experimentally measured F-scan profile \cite{Chalupsky2010}. Results shown in the following figures are taken from the $0.4 \times 0.4$ \SI{}{\micro\meter}$^2$ sample region indicated by the dashed box.}
\label{intensity}
\end{figure}

The experimental setup differed from that of Rackstraw's experiment as described in the previous section only in the XFEL parameters used. The pulse energy was half as much at 0.8--1.4 mJ, the focal spot was slightly larger at $\sim$\SI{9}{\micro\meter}$^2$, and the electron bunch duration was slightly lower at 80 fs. PICLS simulations were performed with parameters matched to those of the associated SCFLY simulations. A \SI{1.4}{\micro\meter} solid Al target is irradiated normally by a 1 mJ x-ray laser pulse with an 80 fs FWHM Gaussian temporal intensity profile and radial profile fit to the measured F-scan profile \cite{Chalupsky2010} as shown in Fig.\,\,\ref{intensity}, resulting in a peak intensity of $1.36\times 10^{17}$ W/cm$^2$. The x-ray laser source has a constant bandwidth of $\sim$3 eV due to the chosen resolution of the photon energy group. The effect of such a small bandwidth should not be significant as the experimental bandwidth was only $\sim$0.4\% ($\sim$7 eV). The initial average free-electron kinetic energy is set to 7 eV, which corresponds to the mean energy of the Fermi-Dirac distribution ($\tfrac{3}{5}E_{\text{F}}$) for aluminum.

Figure \ref{intensity} shows the spatially resolved electron energy density resulting from an x-ray laser pulse with 1700 eV photons at four times during the interaction to illustrate the spatial and temporal variation of the plasma. Though the central plasma along the laser axis becomes homogeneous by the end of the interaction, a longitudinal gradient can be seen on axis for early times and off axis for all times since the photoionization rate is large for $h\nu$ near the absorption edge. In the present study we are not concerned with effects of spatial and temporal gradients, and the following results are peak values averaged over a small $0.4 \times 0.4$ \SI{}{\micro\meter}$^2$ sample region just inside the target surface along the beam axis (as indicated in the figure) within which the plasma conditions are approximately uniform.

\begin{figure}[htbp]
\includegraphics[width=8.5cm]{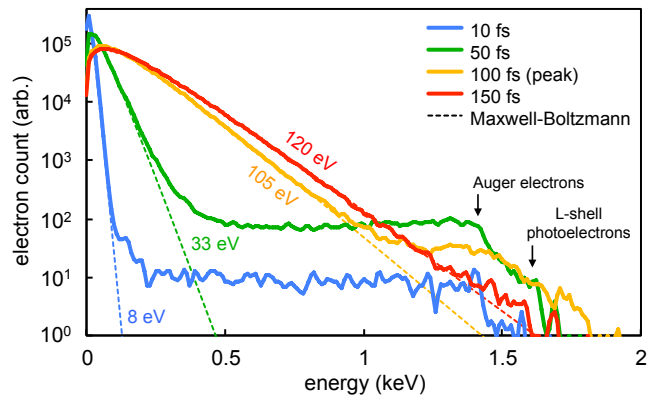}
\caption{(Color online) Electron energy distribution at four times during the interaction resulting from an x-ray laser pulse with 1700 eV photons. The dashed lines indicate a Maxwellian distribution fit to the thermalized component of the distribution to emphasize the non-thermal Auger electrons.}
\label{distribution}
\end{figure}

As discussed in Section \ref{sec-colionrec}, the creation of a plasma by intense x-ray laser radiation is a distinctly non-thermal process. In aluminum, the plasma is seeded mainly by energetic $\sim$1.4 keV KLL Auger electrons and photoelectrons with energies of $h\nu-E_{\text{K}}$, where $E_{\text{K}}$ is the K-shell binding energy. The defining quality of the particle-in-cell technique is that it is capable of supporting virtually any particle distribution and can easily incorporate the process of thermalization with collision models. Figure \ref{distribution} shows the electron energy distribution resulting from a simulation with $h\nu=1700$ eV at different times during the interaction. Also shown are Maxwell-Boltzmann fits to the thermalized component of the distribution (dashed lines) and the corresponding temperatures. The initially $\sim$1.4 keV Auger electrons gradually lose energy as they thermalize via collisional impact ionizations and binary collisions. Since the photoelectrons have energies $<150$ eV, they are not distinguishable from the bulk electrons. The fraction of non-thermal electrons is small but can carry a significant portion of the total energy. For example, at 50 fs the fraction of electrons with kinetic energy above 500 eV accounts for only 0.7\% of the free electron population but contains 11.1\% of the total free electron energy. By 100 fs at peak x-ray intensity, the Auger ionization rate is rapidly decreasing and the Auger electron population is increasingly insignificant compared to the thermalized bulk population.

\begin{figure}[htbp]
\includegraphics[width=7.5cm]{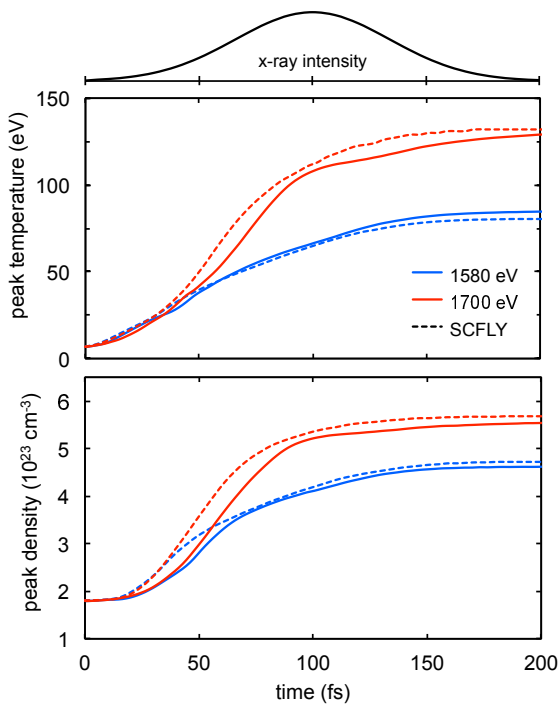}
\caption{(Color online) Evolution of the peak (on-axis) $T_e$ (top) and $n_e$ (bottom) for PICLS (solid curves) and SCFLY (dashed curves) simulations with $h\nu=1580$ eV and 1700 eV. The corresponding time-dependent x-ray intensity is shown above the figures.}
\label{Te_ne_vs_t}
\end{figure}

Further insight into the plasma creation can be gained by examining the time evolution of the $T_e$, $n_e$, and the ion charge state distribution (CSD). Figure \ref{Te_ne_vs_t} shows $T_e(t)$ and $n_e(t)$ resulting from PICLS and SCFLY simulations with $h\nu=1580$ eV and 1700 eV. The total absorption and thus heating and ionization increases with $h\nu$ because higher energy photons can photoionize the Al ions to higher charge states before the increasing K-edge energy surpasses the XFEL $h\nu$ preventing further absorption. For example, pulses with 1580 eV, 1700 eV, and 1820 eV photons can approximately photoionize up to Al$^{5+}$, Al$^{8+}$, and Al$^{11+}$, respectively. Note that the PICLS $T_e$ shown in the figure is not the $T_e$ of the thermalized component of the distribution, but is estimated by assuming an ideal gas as $k_{\text{B}}T_e=(2/3)E_{\text{ave}}$, where $E_{\text{ave}}$ is the average kinetic energy of all free electrons. Since the total absorption is similar between the two codes as demonstrated in the previous section, and since the SCFLY EOS is used in the PICLS recombination model to constrain $\bar{Z}$ as discussed in Sec. \ref{sec-colionrec}, the final values of $T_e$ and $n_e$ after the interaction are similar. The primary difference between the results is a slight delay of $\sim$10 fs resulting from a difference in the collisional ionization rates.

\begin{figure}[htbp]
\includegraphics[width=8.0cm]{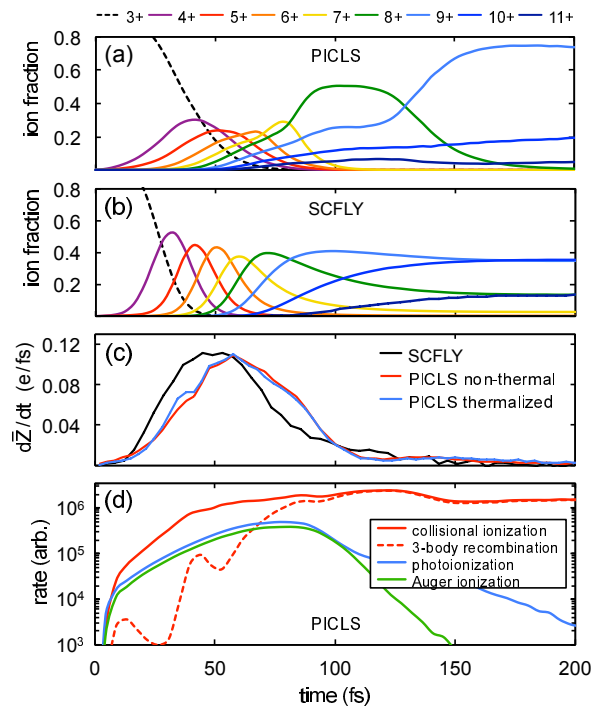}
\caption{(Color online) Evolution of the ion CSD from PICLS (a) and SCFLY (b) simulations with $h\nu=1700$ eV. (c) Comparison of the corresponding rates of change of the average ion charge between the two codes as well as for the PICLS case of forced thermalization. (d) Ionization rates and 3-body recombination rate for the non-thermal PICLS simulation.}
\label{CSD-rates}
\end{figure}

The corresponding evolution of the ion fractional population from the PICLS and SCFLY simulations with $h\nu=1700$ eV are shown in Fig.\,\,\ref{CSD-rates}(a) and (b), respectively. In contrast to the SCFLY simulation in which the CSD tends to progress through charge states sequentially such that only about 3 charge states are significantly populated at any given time, the PICLS simulation tends to have a more spread out CSD resulting from a slightly lower collisional ionization rate. The rate of change of the average ion charge $d\bar{Z}/dt$ from the two simulations is shown in Fig.\,\,\ref{CSD-rates}(c), where it can be seen that the approximate overall effect of the differences between the two codes is a shift of about 10 fs as was observed in the $T_e$ and $n_e$ evolution. The collisional ionization rate quickly surpasses the photoionization rate after only a few femtoseconds and thus dominates the CSD evolution as seen in Fig.\,\,\ref{CSD-rates}(d). The non-physical oscillations in the recombination rate result from the on-off nature of the model as discussed in Sec. \ref{sec-colionrec}.

\subsection{Effects of non-thermal electrons}

CR codes and other codes used to simulate XFEL-driven plasmas typically assume a Maxwell-Boltzmann particle distribution and thus instantaneous thermalization of the source of energetic photoelectrons and Auger electrons. However, since the collisional ionization cross section depends on the energy of the impact electron (see Fig.\,\,\ref{fig-colionrec}(a)), if a significant fraction of the absorbed energy is carried by a relatively small number of high energy electrons with reduced ionization cross sections, then the initial ionization rate can be overestimated if that energy is assumed to be distributed among a much greater number of lower energy electrons according to a thermalized distribution. It is therefore important to understand what the effects of the non-thermal electrons are and under what circumstances they may be neglected.

In Ref. \cite{Ciricosta2016}, the $T_e$ and $n_e$ resulting from an SCFLY simulation similar to those discussed in the previous section were found to have no deviation from those in a modified simulation in which the Auger electron distribution was treated separately from the bulk free electron distribution in an attempt to prevent instantaneous thermalization of the energetic Auger electrons. However, the assumption was made that an Auger electron would become thermalized upon first collision, which is not generally the case since it might take several collisions to completely thermalize, producing secondary, non-thermal, collisionally-ionized electrons in the process. It would be beneficial then to re-examine the problem in a more self-consistent manner. 

In this section, we investigate the effects of the non-thermal electrons by comparing the results of PICLS simulations against identical simulations in which the energetic Auger electrons and photoelectrons are forced to instantly thermalize. To achieve instant thermalization, immediately after the photoionization and Auger ionization processes have been performed (once every $\Delta t_{\text{rad}}$), the cell temperature is calculated assuming an ideal gas as $k_{\text{B}}T_e=(2/3)E_{\text{ave}}$, where $E_{\text{ave}}$ is the average kinetic energy of the free electrons including the non-thermal photoelectrons and Auger electrons. The free electron energies are then randomly assigned following a Maxwell-Boltzmann distribution with temperature $k_{\text{B}}T_e$ by choosing the momentum components of each electron ($p_x$, $p_y$, $p_z$) randomly from a normal distribution with zero mean and standard deviation of 1. The small difference in cell energy resulting from the randomized process is corrected for to ensure conservation of energy and momentum.

PICLS simulations were performed with parameters identical to the simulations in the previous section with $h\nu=1700$ eV, both with and without the forced thermalization process. As seen in Fig.\,\,\ref{CSD-rates}(c), the resulting evolutions of $d\bar{Z}/dt$ are nearly identical between the two cases. Similarly, the evolutions of $T_e$, $n_e$, and CSD are nearly identical. Before the pulse peak, the fraction of the absorbed energy carried by non-thermal electrons is less than 10\%, which is somewhat small but not insignificant, and we expect to see a difference in the ionization rate. However, the collisional ionization cross section for the source 1.4 keV Auger electrons is actually higher than the cross sections for the vast majority of thermalized electrons at early times so that assuming a Maxwellian distribution actually \textit{reduces} the ionization rate initially. Indeed, close inspection of Fig.\,\,\ref{CSD-rates}(c) shows a slightly lower ionization rate in the thermalized case for the first 20 fs, at which point the temperature is high enough that the ionization rate of the thermalized distribution becomes higher. Thus the thermalized simulation first underestimates and then overestimates the collisional ionization rate such that the net effect is approximately cancelled, and the plasma properties are unchanged. 

We have shown that, for the conditions of the particular experiment under consideration, the effect of the non-thermal Auger electrons is negligible. However, it remains to be shown to what extent this result will be true. In order to better understand the limits on the validity of the assumption of instantaneous thermalization, we examine a more extreme case in which a solid aluminum target is irradiated by an intense x-ray laser pulse with 10 keV photons. In addition to the $\sim$1.4 keV Auger electrons, the 10 keV x-rays will produce an even larger number of $\sim$8 keV photoelectrons in contrast to the $\sim$0.15 keV photoelectrons in the previous simulations. Since the K-shell photoionization cross section for a 10 keV photon is about two orders of magnitude lower than that for a photon with energy near the K-edge (as seen in Fig.\,\,\ref{photo-auger}(a)), the peak intensity is increased to $10^{19}$ W/cm$^2$ by increasing the pulse energy to 3.8 mJ, decreasing the pulse duration to 20 fs FWHM, and decreasing the beam radius to \SI{1}{\micro\meter} FWHM. The radial profile is a super-Gaussian that is close to a square profile, enabling a larger particle sample region. The remaining parameters are similar to those of the previous simulations.

\begin{figure}[htbp]
\includegraphics[width=8.5cm]{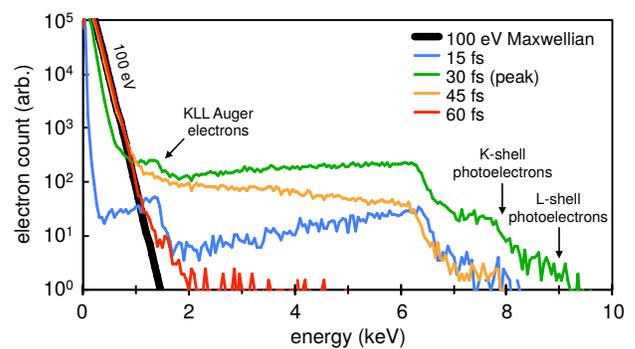}
\caption{(Color online) Electron distribution resulting from a solid Al plasma driven by an x-ray laser pulse with 10 keV photons at different times. The thermalized 100 eV Maxwell-Boltzmann distribution is shown in black.}
\label{fig-dist-10keV}
\end{figure}

The resulting highly non-thermal electron distribution can be seen at different times during the interaction in Fig.\,\,\ref{fig-dist-10keV}. In contrast to the previous simulations, the majority of the plasma electron energy can be carried by the energetic photoelectrons and Auger electrons. For example, at the peak of the x-ray pulse (30 fs), 56\% of the total free electron energy is carried by electrons with energies above 700 eV though they account for only 3.6\% of the population. By the end of the pulse at 60 fs, the plasma has thermalized to a temperature of $\sim$100 eV.

\begin{figure}[htbp]
\includegraphics[width=7.5cm]{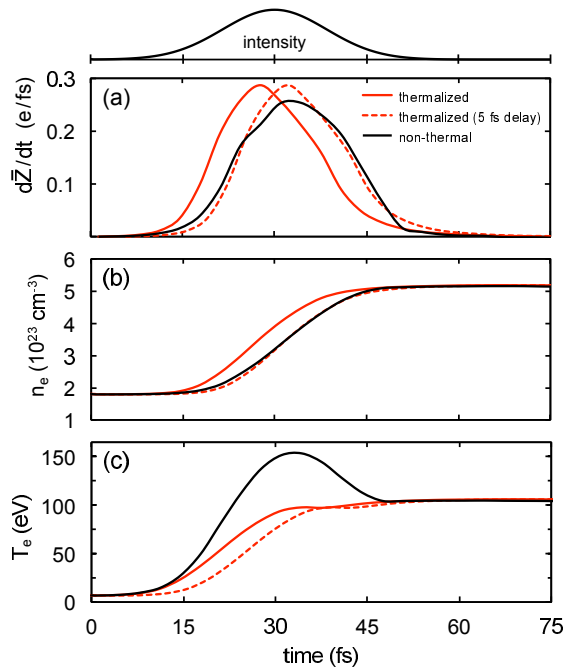}
\caption{(Color online) Evolution of the rate of change of the average ion charge (a), electron density (b), and electron temperature (c) resulting from PICLS simulations of a solid Al plasma driven by 10 keV photons, both with (red) and without (black) forced instantaneous thermalization. The dashed curves are the results from the thermalized case shifted in time by 5 fs.}
\label{fig-evo-10keV}
\end{figure}

The rate of change of the average ion charge $d\bar{Z}/dt$ can be seen in Fig.\,\,\ref{fig-evo-10keV}(a) for the simulations with and without forced thermalization. As expected, the ionization rate in the thermalized simulation is initially overestimated. A faster progression though increasing ion charge states would mean that predicted levels of K$_{\alpha}$ emission from lower charge states would be somewhat underestimated by assuming instant thermalization. However, the progression rate through charge states is roughly the same. Instead, the net effect of the non-thermal distribution is approximately a shift in time of about 5 fs as shown by the dashed curve in the figure.
The corresponding $n_e$ ($\propto \bar{Z}$) is shown in Fig.\,\,\ref{fig-evo-10keV}(b), where it is seen that the $n_e$ resulting from the non-thermal distribution is nearly identical to that from the thermalized distribution delayed by 5 fs. Fig.\,\,\ref{fig-evo-10keV}(c) shows the evolution of $T_e$. The peak $T_e$ in the non-thermal simulation becomes much larger than that in the thermalized simulation before returning to the same level. This happens because the temperature is calculated from the average kinetic energy of all free electrons, including non-thermal electrons, and the extra energy is eventually lost to collisional ionizations as the energetic electrons thermalize. The total x-ray absorption does not change since the final $n_e$ and $T_e$ are identical between the two simulations.

Experimental measurements typically consist of time-integrated measurements of the x-ray absorption and emission of the plasma. If the effect of the energetic electrons is only to shift the ionization rate in time by a few femtoseconds, then simulated results of time-integrated properties will be unaffected. Thus the assumption of a Maxwellian distribution seems reasonable when simulating such time-integrated properties of XFEL-driven solid-density plasmas for configurations similar to those in this study.

\section{Summary \label{sec-summary}}

A particle-in-cell code, PICLS, which self-consistently solves for the radiation transport, has been extended with Monte-Carlo models for photoionization and the resulting decay processes of K-shell vacancies, including KLL Auger ionization, enabling the fully kinetic simulation of solid-density XFEL-driven plasmas. We have discussed in detail the algorithms used for these models as well as for the models of radiation transport, collisional ionization, three-body recombination, and continuum-lowering. As an initial test of the newly-developed models, we simulated two LCLS experiments investigating properties of solid-density aluminum HED plasmas driven by K-shell photoionization from intense XFEL pulses with photon energies scanned across the absorption K-edge \cite{Vinko2012,Rackstraw2015}. The pulse transmissions resulting from PICLS simulations using the Ecker-Kr{\"o}ll continuum-lowering model were found to agree well with the experimentally measured values, while those from simulations using the Stewart-Pyatt model did not agree, as the predicted K-edge energies were far too high and the total absorption was too low. Additionally, the time-resolved electron temperature, density, and ion charge state distribution from PICLS simulations were found to agree well with those values from simulations by the collisional-radiative (CR) code SCFLY, which has been used to accurately reproduce the experimental time-integrated K$_\alpha$ emission spectra with excellent accuracy \cite{Ciricosta2016,Ciricosta2012,Vinko2015}.

XFEL-driven plasmas are seeded by non-thermal photoelectrons and Auger electrons that ionize and heat the plasma through collisions. CR codes and other codes used to simulate such plasmas typically assume a Maxwellian particle distribution and thus instantaneous thermalization of the source of energetic electrons. However, since the collisional ionization cross section depends on the energy of the impact electron, the simulated ionization rate and related plasma properties may be affected by assuming a thermalized distribution. As PICLS is fundamentally a particle-in-cell code, it is able to account for non-Maxwellian particle distributions as well as the thermalization process via Monte-Carlo collision models. The effects of the non-thermal electrons were investigated by comparing PICLS simulation results with results of identical simulations in which the non-thermal electrons were forced to instantly thermalize. For the conditions of the particular experiments under consideration in which the plasmas were seeded by $\sim$1.4 keV Auger electrons, the difference in ionization rate is negligible. Additional simulations were performed for the more extreme case of a plasma driven by 10 keV photons, which produces a distribution seeded mainly by 8 keV photoelectrons. It was observed that, by forcing the electrons to thermalize, the initial ionization rate was overestimated such that the overall effect was approximately a shift in time of the ionization process by several femtoseconds. Such an effect should not matter when simulating time-integrated plasma properties such as K$_{\alpha}$ emission spectra or total pulse transmission. 

Future applications of PICLS will focus on experimental regimes in which the capabilities of a particle-in-cell code might offer more insight than other modeling techniques. For example, XFEL-driven plasmas in lower density targets will require longer thermalization times, and the assumption of instantaneous thermalization may no longer be valid. Additionally, plasmas created by highly focused XFEL pulses with sub-micron spots and peak intensities approaching $10^{20}$ W/cm$^2$ can create keV plasmas in higher-$Z$ targets. In this regime, radial energy transport and induced electric fields can play an important role in determining the temperature and density of the plasma. Future development efforts will be directed towards the addition of new physics models or improvement of current models to address limitations of the code. This includes the addition of models for resonant bound-bound transitions and higher-order Auger processes to more accurately describe interactions with higher-$Z$ materials, as well as improvement of the quality of the K$_{\alpha}$ emission data to enable the generation of spectroscopic-quality synthetic spectra which can be compared directly to measurements. Additionally, as discussed in Sec. \ref{sec-colionrec}, the three-body recombination model needs to be redesigned so that it does not depend on a chosen equation of state but is instead based on cross sections so that it is analogous and complementary to the impact ionization model.

\begin{acknowledgments}
R.R., Y.S., and R.C.M. would like to thank S. Vinko and D. Rackstraw for providing data for use in the figures, and we also thank S. Vinko, B. Nagler, and H.J. Lee for the fruitful discussions and collaboration. We gratefully acknowledge support by the US DOE-OFES under Contract No. DE-SC0008827 and JSPS KAKENHI Grant No. JP15K21767. T.J. acknowledges support by the Joint Institute for Fusion Theory (JIFT), Japan/US Fusion Research Collaboration.
\end{acknowledgments}


\end{document}